\documentclass{article}
\usepackage[a4paper=true,hypertex]{hyperref}
\usepackage{epsfig}
\author{Silvia Mollerach and Esteban Roulet\\
{\it CONICET, Centro At\'omico Bariloche,}\\
{\it Av. Bustillo 9500 (8400) Argentina.}}

\title{Magnetic diffusion effects on the Ultra-High Energy \\
 Cosmic Ray spectrum and composition}

\begin{document}
\maketitle

\begin{abstract}
 We discuss the effects of diffusion of high energy cosmic rays in turbulent extra-galactic magnetic fields. We find an approximate expression for the 
 low energy suppression of the spectrum of the different mass components (with charge $Z$)  in the case in which this suppression happens at energies below $\sim Z$~EeV, so that energy losses are dominated by the adiabatic ones.
 The low energy suppression appears when cosmic rays from the closest sources take a time comparable to the age of the Universe to reach the Earth. This occurs for energies $E< Z\, {\rm EeV}\,(B/{{\rm nG}})\sqrt{ l_c/{\rm Mpc}}(d_s/70\ {\rm Mpc})$ in terms of the magnetic field RMS strength  $B$, its coherence length $l_c$ and the typical separation between sources $d_s$. 
 We apply this to  scenarios in which the sources produce a mixed composition and have a relatively low maximum rigidity ($E_{max}\sim (2$--$10) Z  $~EeV), finding that diffusion  has a significant effect on the resulting spectrum, the average mass and on its spread, in particular reducing this last one. 
For reasonable values of $B$ and $l_c$ these effects can help to reproduce the composition trends observed by the Auger Collaboration for source spectra compatible with Fermi acceleration.
\end{abstract}

\section{Introduction}
Charged particles traversing a region with a turbulent magnetic field characterized by an RMS strength $B$ can be in three different propagation regimes depending on the relation between the magnetic field coherence length $l_c$, the source distance $r_s$ and the  Larmor radius
\begin{equation}
r_L=\frac{E}{ZeB}\simeq 1.1 \frac{E_{\rm EeV}}{ZB_{{\rm nG}}} \ {\rm Mpc},
\end{equation}
with  $Ze$ the particle charge, $B_{{\rm nG}}\equiv B/{\rm nG}$ and $E_{\rm EeV}\equiv E/{\rm EeV}$ (with ${\rm EeV}=10^{18}$~eV). 

Defining the critical energy $E_c$ such that  $r_L(E_c)=l_c$, one obtains 
 $E_c\simeq 0.9 Z B_{{\rm nG}}(l_c/{\rm Mpc})$~EeV. 
For energies $E<E_c$,  one has that $r_L<l_c$ so that  diffusion occurs
through the resonant  scattering of the  particles on the magnetic field components with wavelengths of order $r_L$. This propagation regime is usually referred to as the quasi-linear regime or resonant diffusion. The diffusion length $l_D$  depends on the power of the magnetic field on the scales of the Larmor radius, and hence depends on the turbulent magnetic field spectrum.  Typically one has  $l_D\simeq l_c(E/E_c)^\alpha$, where for a Kolmogorov spectrum  $\alpha=1/3$, for a Kraichnan spectrum $\alpha=1/2$ while the so-called Bohm diffusion adopts $\alpha=1$ (corresponding to the assumption $l_D=r_L$). 
The diffusion length $l_D$ corresponds to the distance for which the typical particle deflection is 1~rad, and hence is related to the basic step in the random walk of the diffusing particles. The diffusion coefficient is just $D=c l_D/3$ and the rectilinear distance $\Delta r$ travelled by particles in their random walk after a time $\Delta t$ satisfies $\langle (\Delta r)^2\rangle = 6D\Delta t$.

When  $E>E_c$ (i.e. for $r_L>l_c$) the scattering is non-resonant and is the result of many small deflections, with $\delta\theta\simeq l_c/r_L$ in each coherence length $l_c$. As a result, after $N\simeq l_D/l_c$ steps the deflection is $\Delta\theta\simeq \sqrt{N}\delta\theta=1$~rad, so that  one obtains $l_D\simeq l_c(E/E_c)^2$ and we see that the diffusion length strongly increases with energy. Two regimes are possible here depending on the overall distance travelled (i.e. depending on the source distance $r_s$). As long as $l_D\ll r_s$ spatial diffusion of the particles takes place, while at energies for which $l_D$ becomes larger than $r_s$ the overall deflections become small and the quasi-rectilinear propagation of the particles takes place, leading to just a small angular diffusion of the particle trajectories. 

 Cosmic rays (CRs) arriving to the Earth are deflected by both the galactic and the extra-galactic magnetic fields.  Regarding the deflections in the galactic fields, which have typical strengths of few $\mu$G with the random component having  $l_c\simeq  10$--100~pc, the critical energy is  $E_c\simeq 0.06Z(B/3\,\mu {\rm G})(l_c/20\,{\rm pc})$~EeV and the particles reach the quasi-rectilinear propagation in the turbulent field ($l_D\simeq 1$~kpc, which is the typical vertical extent of the random magnetic field) for energies $E\simeq \sqrt{{\rm kpc}/l_c}E_c\simeq few Z\,10^{17}$~eV, and somewhat larger for directions along the galactic plane. Actually, the dominant deflections in the galactic magnetic field are expected to be those caused by the regular component, due to its larger coherence length of order $\sim 1$~kpc. As a result, the galactic magnetic field deflections at EeV energies are large even for CR protons. Although these deflections can significantly modify the arrival directions (hence diluting possible anisotropy signals) they do not affect the expected overall spectrum nor the composition of the extra-galactic cosmic rays.

Focusing now on the extra-galactic magnetic fields, the simplest assumption is to consider a uniform turbulent field permeating intergalactic space. Strengths for $B$ in the range from 0.01~nG  up to 100~nG and coherence lengths $l_c$ from 10~kpc (as could be associated to galactic outflows) up to 50~Mpc (as could be associated to supercluster scales) have been considered\footnote{Notice that for a Kolmogorov spectrum with maximum scale of turbulence $l_{max}$ and minimum one $l_{min}\ll l_{max}$ one has that $l_c\simeq l_{max}/5$ \cite{ha02}. 
}. For instance the simulations in \cite{si04} obtain $B\sim 100$~nG in filaments while  $B\sim (10^{-3}$--100)~nG outside filaments,  while ref.~\cite{do04} obtains values $B< 0.01$~nG in voids and $B\sim 0.1$~nG in filaments, with fields inside clusters being at the $\mu$G level.
The possible impact on the CR diffusion of non-uniform magnetic fields and of its possible cosmological evolution  were discussed in \cite{ko08}.  A review with some magnetic field upper bounds obtained in the literature for different values of $l_c$ can be found in \cite{wi03}.

The average separation between the UHECR sources $d_s$ is related to their density $n_s$ through $d_s\simeq n_s^{-1/3}$, with $d_s\simeq 10$~Mpc for $n_s=10^{-3}$~Mpc$^{-3}$ while $d_s\simeq 100$~Mpc for $n_s=10^{-6}$~Mpc$^{-3}$, which covers the range of densities usually considered. CRs from a source at distance $d_s$ will have a diffusive propagation  (i.e. $l_D<d_s$) as long as $E<E_s$, where (assuming the non-resonant diffusive regime to hold) one gets 
\begin{equation}
E_s\simeq 3 Z B_{{\rm nG}}\sqrt{\frac{l_cd_s}{10\,{\rm Mpc}^2}}\, {\rm EeV}.
\label{es.eq}
\end{equation}
Hence, even protons from the closest sources may have a diffusive behavior up to EeV energies for  values of $B\sqrt{l_c}\sim {\rm nG}\sqrt{{\rm Mpc}}$, and up to higher energies for sources farther away, with nuclei having a similar behavior but shifted up in energy by a factor $Z$ (i.e. equal for the same rigidities).
 
Two additional ingredients are fundamental in shaping the CR spectra of the different elements: the interactions with the radiation backgrounds and cosmological evolution effects.
CR energy losses arise from adiabatic losses (d$E/{\rm d}t=-HE$, with $H=\dot a/a$ being Hubble's constant in terms of the scale factor $a$), which are present at all energies, and from the interaction losses due to different processes. The pair ($e^{+}e^{-})$ creation losses are relevant for interactions with CMB photons of energy $\varepsilon\sim 10^{-3}$~eV, requiring CR Lorentz factors $\gamma>10^9$ and hence $E>A$~EeV for nuclei with mass number $A$. Actually, the energy loss length for protons becomes  smaller than the Hubble horizon $R_H=c/H_0$, with $H_0\simeq 70$~km/s/Mpc, for energies $E_p>2$~EeV, while for heavy nuclei the combination of increased threshold and larger pair production cross section, $\sigma \propto Z^2$, leads to an energy loss length comparable to $R_H$ for $E_A\simeq A$~EeV \cite{al11}. The photo-pion production off CMB photons becomes significant only for very high energies, $E/A>50$~EeV, while the photo-disintegration of nuclei is relevant for $E>2A$~EeV for interactions with  CMB photons and down to lower energies for interactions with higher energy background photons such as IR or optical ones. However, for interactions with IR photons the energy loss length for photo-disintegration is larger than the Hubble horizon except for heavy nuclei such as Fe for which it is larger than $R_H$ only for $E<30$~ EeV \cite{al11}. As a result of these considerations, we can say that in general for $E<Z$~EeV the energy losses are mainly due to the adiabatic ones while the ones due to interactions are small.

In addition to the adiabatic losses, cosmological effects also enter in the increase of the CMB density and temperature with redshift, in possible redshift evolution of the other radiation backgrounds, of the CR source density and emissivity and eventually also in the magnetic field evolution.

\section{The low energy spectral suppression}

The effects of CR diffusion in extra-galactic magnetic fields were discussed in several works in the past (see e.g. \cite{ac99,bl99,st00,yo03,le05,be06,be07,si07,gl08,ko08,ta11}. A detailed analytic treatment of the proton diffusion  in an expanding Universe was performed by Berezinsky and Gazizov \cite{be06,be07}, generalizing the Syrovatskii solution \cite{sy59}  which provides the CR density as a function of the distance to a source  in a static Universe and accounting for the energy losses during propagation. 

The general solution obtained in \cite{be06} is that the contribution to the flux from a given source at comoving distance $r_s$ (i.e. corresponding to its present distance) is
\begin{equation}
J_s(E)=\frac{c}{4\pi}\int_0^{z_{max}}{\rm d}z\, \left|\frac{{\rm d}t}{{\rm d}z}\right| Q[E_g(E,z),z)]\frac{\exp[-r_s^2/(4\lambda^2)]}{(4\pi\lambda^2)^{3/2}} \frac{dE_g}{dE},
\label{js.eq}
\end{equation}
where $z_{max}$ is the maximum source redshift, $E_g(E,z)$ is the original energy at redshift $z$ of a particle having energy $E$ at redshift $z=0$, $Q$ is the source spectra, which we will take as a sum of the contributions $Q_Z$ from the different charges, and we will adopt  a power law with a maximum rigidity cutoff $Z E_{max}$, modeled as $Q_Z(E,z)=\xi_Zf(z)E^{-\gamma}/{\rm cosh}(E/ZE_{max})$. The  parameter $\xi_Z$ accounts for the relative contribution of nuclei of charge $Z$ to the flux, the function $f(z)$ accounts for the eventual redshift evolution of the source emissivity and the spectral slope $\gamma$ is expected to be in the range 2--2.4 in Fermi type diffusive shock acceleration. The cosh$^{-1}(E/ZE_{max})$ allows to implement a smooth suppression of the spectrum without affecting significantly the spectral shape at $E<ZE_{max}/2$, unlike what happens when an exponential cutoff is adopted.
The parameter $\lambda$ in eq.~(\ref{js.eq}) is the generalized Syrovatskii variable, given by
\begin{equation}
\lambda^2(E,z)=\int_0^z{\rm d}z'\,\left| \frac{{\rm d}t}{{\rm d}z'}\right| \frac{D(E_g,z')}{a^2(z')},
\label{lambda.eq}
\end{equation}
with $a(z)=1/(1+z)$ and $D$ the diffusion coefficient. Moreover
\begin{equation}
\left| \frac{{\rm d}t}{{\rm d}z}\right|=\frac{1}{H_0(1+z)\sqrt{(1+z)^3\Omega_m+\Omega_\Lambda}}, 
\end{equation}
where $H_0\simeq 70$~km/s/Mpc is the present Hubble constant, $\Omega_m\simeq 0.27$ the matter content and $\Omega_\Lambda\simeq 0.73$ the cosmological constant contribution at present.

Note that $\lambda(E,z)$ can be interpreted as the typical radial comoving distance from the source traversed by a particle leaving from it at redshift $z$ and arriving at present with energy $E$. 

Although eq.~(\ref{js.eq}) was derived for protons, it can in principle also be applied to nuclei expressing it in terms of the particle rigidities. A relevant fact of photo-disintegration processes for nuclei is that they approximately conserve the  Lorentz factor and the rigidity of the main fragment and hence  do not affect significantly the diffusion properties of the particles\footnote{An exception is for the secondary protons, which having $Z=A=1$ have approximately half the rigidity of their parent nuclei.}.
A possible complication is that when photo-disintegration losses are important the source term $Q$ should refer to the one of the primary nucleus which gave rise to the observed one, and this is hard to obtain, in particular due to the stochastic nature of the process. However, at low energies when the photo-disintegration losses become small (which is the regime we will focus in below) the expression in eq.~(\ref{js.eq}) can be used reliably.

The general results of the diffusion effects is to suppress the CR flux at low rigidities, since particles are not able to arrive to the observer from distant sources and take a much longer time than in the case of rectilinear propagation to arrive from the nearby ones. However, a very important result is the so-called propagation theorem \cite{al04}, which states that as long as the distance to the nearest sources  is smaller than the other relevant length scales (diffusion length and energy loss length), the total CR flux will be the same as that obtained ignoring magnetic field effects and for a continuous distribution of sources. This means that even at energies for which far away sources do not contribute anymore, as long as the observer lies within the diffusion sphere  of the nearby sources the spectrum is unchanged and only when the nearest sources get suppressed one has that the overall spectrum is modified.  It is hence important to study in detail the suppression effect of the closest sources. 

A crucial relation to prove the propagation theorem is to sum over the sources in eq.~(\ref{js.eq}), assuming for simplicity equal source luminosities,  and in the limit of small source separations replace the sum as $\sum\to n_s \int {\rm d}r\,4\pi r^2$, and use that
\begin{equation}
\int_0^\infty {\rm d}r\,4\pi r^2 \frac{\exp(-r^2/4\lambda^2)}{(4\pi\lambda^2)^{3/2}}=1,
\label{one}
\end{equation}
and hence we see that the diffusion effects do not modify the total flux for a continuous source distribution.

\begin{figure}[t]
\centerline{\epsfig{width=2in,angle=-90,file=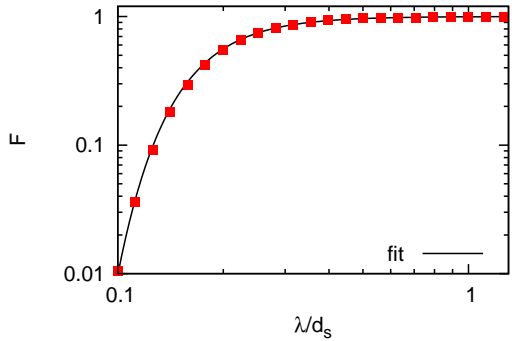}}
\vskip 1.0 truecm
\caption{Suppression factor $F$ as a function of $\lambda/d_s$ and fitting function.} 
\label{f.fig}
\end{figure}

In order to estimate the suppression due to the finite distance to the sources we will compute the actual sum adopting a particular realization of distance distributions corresponding to a uniform source density, taking the source distances from the observer as $r_i=(3/4\pi)^{1/3} d_s \Gamma(i+1/3)/(i-1)!$, which corresponds to the average value of the distance to the $i$-th closest source for  a uniform source distribution\footnote{Different possible realizations of the distance distributions consistent with a uniform density will give rise to slight variations in the suppression factor and lead to a 'cosmic variance' associated to the final result, see e.g. \cite{gl08}.}.
Hence, when one performs the sum over sources with a discrete source distribution, instead of getting unity as in eq.~(\ref{one}) one gets the factor
\begin{equation}
F\equiv \frac{1}{n_s}\sum_i\frac{\exp(-r_i^2/4\lambda^2)}{(4\pi\lambda^2)^{3/2}}.
\label{f.eq}
\end{equation}
The resulting factor is plotted with dots in fig.~\ref{f.fig} as a function of $\lambda/d_s$ and shows a pronounced suppression when $\lambda<d_s/3$. Also shown with solid line is a fit to the results obtained of the form
\begin{equation}
F_{fit}\simeq \exp(-(d_s/6\lambda)^3).
\label{ffit.eq}
\end{equation}

We will be interested in the following in the consideration of scenarios in which the suppression happens for protons at energies not larger than $\sim 1$~EeV (and hence below $Z$~EeV for nuclei of charge $Z$). This is motivated by the fact that there is evidence for a non-negligible proton fraction at EeV energies   from Auger, Hires and TA experiments \cite{augerx,hiresx,tax}.  This light component should presumably be of extra-galactic origin since galactic protons would be in tension with the strict bounds on dipolar anisotropies obtained by Auger in this energy range \cite{augerls12}. Focusing on energies below $Z$~EeV leads to the simplification that the main energy loss process is just the one due to adiabatic losses, and we will obtain in this case a simpler analytic  expression for the suppression of the flux due to the magnetic diffusion from the discrete set of sources. In an Appendix we discuss the effects of including pair production losses in the case of protons to illustrate the validity of our approach.

Regarding the diffusion coefficients, we note that the detailed values of the diffusion length in the presence of turbulence can be obtained by Monte Carlo simulations of the particle propagation in random magnetic fields and studying the decorrelation times (see e.g. \cite{ca02,ca04}). We will here adopt a fit to the results for Kolmogorov turbulence obtained by Globus et al. \cite{gl08}, which is
\begin{equation}
l_D(E)\simeq l_c\left[(2\pi)^{-2/3}\left(\frac{E}{E_c}\right)^{1/3}+\frac{4\pi}{3}\left(\frac{E}{E_c}\right)^2\right].
\label{ld.eq}
\end{equation}
This expression improves the approximate estimates used in the Introduction by numerical factors of order unity, and is valid both for the resonant and non-resonant regimes. Turbulent spectra different from the Kolmogorov one will have a different power of the energy in the first term and also a slightly different numerical coefficient.

Regarding the cosmological evolution of the magnetic field parameters, note that one expects that $l_c(z)=l_c(0)/(1+z)$, while the magnetic field evolution can be parameterized as $B(z)=(1+z)^{2-m}B(0)$, where the $(1+z)^2$ factor arises from flux conservation, and the factor $m$ was introduced in \cite{be07} to account for additional MHD effects (and taken there as $m=1$). The critical energy at redshift $z$  (i.e. such that $r_L(E_c(z))=l_c(z)$) is just $E_c(z)=E_c(1+z)^{1-m}$, and hence is constant for $m=1$.

\begin{figure}[t]
\centerline{\epsfig{width=3in,angle=0,file=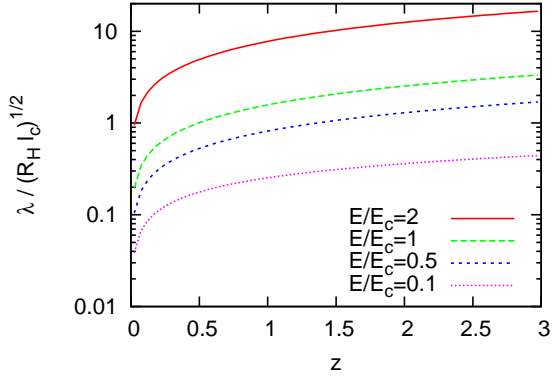}}
\vskip 1.0 truecm
\caption{$\lambda/\sqrt{R_hl_c}$ vs. redshift for different values of $E/E_c$.} 
\label{fig2}
\end{figure}

Putting all this together one gets from eq.~(\ref{lambda.eq}) that
\begin{equation}
\lambda^2(E,z)\simeq \frac{c}{3H_0}\int_0^{z}{\rm d}z'\, \frac{1+z'}{\sqrt{\Omega_m(1+z')^3+\Omega_\Lambda}}\, l_D((1+z')E), 
\end{equation}
with $l_D$ from eq.~(\ref{ld.eq}) expressed in terms of $E_c(z)$ and $l_c(z)$. One then gets
\begin{equation}
\lambda^2(E,z)\simeq \frac{R_H l_c}{3}\int_0^{z}\frac{{\rm d}z'}{\sqrt{\Omega_m(1+z')^3+\Omega_\Lambda}}\,\left[(2\pi)^{-2/3} \left(\frac{(1+z')^mE}{E_c}\right)^{1/3} +\frac{4\pi}{3}\left(\frac{(1+z')^mE}{E_c}\right)^2 \right],
\label{lambda2.eq}
\end{equation}
with $l_c=l_c(0)$ and $R_H=c/H_0$ the Hubble radius.
 The resulting value of $\lambda(E,z)/\sqrt{R_H l_c}$ is plotted in fig.~\ref{fig2} as a function of $z$ and for different values of $E/E_c$,  adopting  $m=1$ for definiteness. Note that $\sqrt{R_Hl_c}\simeq 65$~Mpc$\sqrt{l_c/{\rm Mpc}}$.

One can now combine eqs.~(\ref{f.eq}) and (\ref{lambda2.eq}) with eq.~(\ref{js.eq}) to obtain, after summation over all sources, that the flux from nuclei of charge $Z$ is
\begin{equation}
J_Z(E)\simeq \frac{R_H n_s\xi_Z}{4\pi}\int_0^{z_{max}}\frac{{\rm d}z}{\sqrt{\Omega_m(1+z)^3+\Omega_\Lambda}}f(z)\frac{[E(1+z)]^{-\gamma}}{{\rm cosh}(E(1+z)/ZE_{max})} F\left(\frac{\lambda}{d_s}\right).
\label{jza.eq}
\end{equation}
One can hence define the final suppression factor of the fluxes as
\begin{equation}
G(E/E_c)\equiv \frac{J_Z(E)}{J_Z(E)|_{F=1}},
\end{equation}
i.e. as the ratio of the actual flux to the one that would be obtained for a continuous source distribution (corresponding to $F=1$ in eq.~(\ref{jza.eq})).
 For this computation we can take $E_{max}\gg E_c$, since we will be interested in scenarios for which the suppression happens for energies well below $E_{max}$,  and the results obtained are hence independent of $E_{max}$.
 Note that  once written in terms of $E/E_c$, the suppression factor $G$ is the same for all nuclei.

\begin{figure}[t]
\centerline{\epsfig{width=3in,angle=0,file=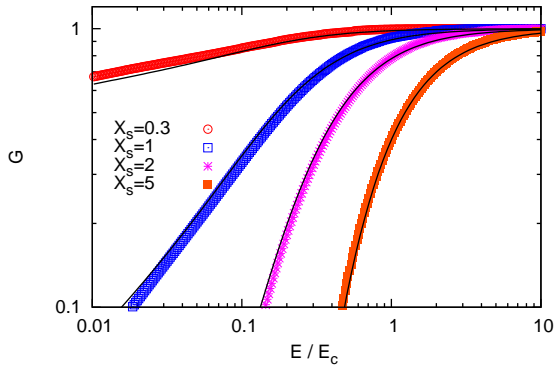}}
\vskip 1.0 truecm
\caption{Suppression factor $G$ vs $E/E_c$ for different values of $X_s$ (with dots) and fitting functions (solid lines).} 
\label{fig3}
\end{figure}

The suppression factor is shown in fig.~\ref{fig3}  adopting $f(z)=const$, $\gamma=2$  and $z_{max}=2$. The resulting values depend on  the average distance between sources $d_s$ and the coherence length $l_c$ through the combination 
\begin{equation}
X_s\equiv \frac{d_s}{\sqrt{R_H l_c}}\simeq \frac{d_s}{65\ \rm Mpc}\sqrt{\frac{\rm Mpc}{l_c}},
\label{xs.eq}
\end{equation}
 and results are displayed for $X_s=0.3$, 1, 2 and 5 as a function of $E/E_c$. 
Approximate fits to these results can be obtained through the expression
 \begin{equation}
G(x)=\exp\left[-\frac{(a\,X_s)^\alpha}{x^\alpha+b x^\beta}\right],
\label{gfit.eq}
\end{equation}
 with $\alpha=1.43$, $\beta=0.19$, $a=0.2$ and $b=0.09$. These fits are also shown in figure~3 with solid lines.
Note that for $E<0.2 E_c$ the diffusion coefficient has the moderate energy dependence $D\sim E^{1/3}$ for the assumed Kolmogorov turbulence and hence the suppression factor decreases more slowly with decreasing energies, while for $E>0.2 E_c$ there is a strong energy dependence of  the suppression factor and for different values of $X_s$ it is essentially just shifted in energy proportionally to $X_s$.

The flux suppression is mainly due to the fact that  at low energies CRs are unable to reach the Earth from the faraway sources and moreover  they take a very long time to arrive to the Earth from the nearby ones, so that the low redshift contribution from these last is absent (in the diffusive regime  the redshift gives a measure of propagation time rather than distance). 
Hence, the suppression factor $G$  has some sensitivity to source evolution effects because these determine the relative contribution to the flux arising from different redshifts. In fig.~4 we show the factor $G$ for $X_s=3$ and different assumptions on the cosmological parameters: the cases of source emissivity evolution as $f(z)\propto (1+z)^a$ with $a=3$ (while $a=0$ in the remaining cases), the case $z_{max}=3$ and that for the magnetic field evolution parameter $m=0$, besides the case corresponding to the parameters adopted in the previous figure. The results vary slightly, except if a very strong source emissivity evolution is adopted, what could lead to a reduced  suppression (larger $G$) due to the larger contribution to the observed CRs from high-redshift (i.e. longer propagation times, allowing for more diffusion). 
In this case
the suppression factor $G$ is similar in shape but gets shifted to lower values of $E/E_c$. Hence, adopting a value for $X_s$ one finds that a given suppression factor would lead to inferred values of $E_c$ larger by a factor about 1.5 in the case of strong evolution with respect to what would be obtained in the case of  uniform source evolution.
 We note that the evolution of the source emissivity considered here is not equivalent to
assuming an evolution in the source density, since increasing this last would decrease the typical source distance, shifting the magnetic suppression effect also to lower energies but due to different physical reasons.

\begin{figure}[t]
\centerline{\epsfig{width=3in,angle=0,file=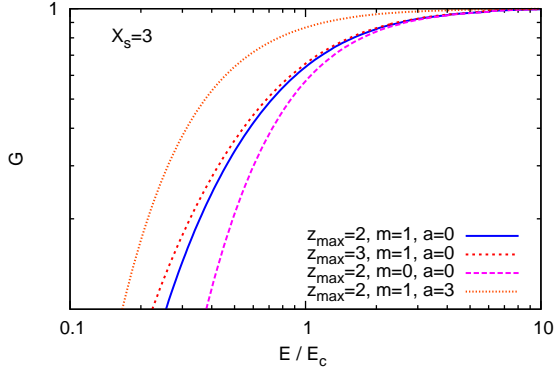}}
\vskip 1.0 truecm
\caption{Suppression factor $G$ vs $E/E_c$ for $X_s=3$ and different assumptions on the cosmological evolution parameters (see text).} 
\label{fig4}
\end{figure}

A necessary condition to have diffusion from the nearest source is that $l_D<d_s$, and hence $E<E_s$ (with $E_s$ in eq.~(\ref{es.eq}), from which one gets $E_s\simeq X_sE_c\sqrt{R_H/d_s}\gg X_sE_c$). Moreover, for the flux from the nearest source to arrive in a typical time scale of the order of the Hubble time, so that the suppression is significant, one needs that $d_s^2> l_D R_H$ and hence one needs, in the regime in which $l_D\simeq  l_c(E/E_c)^2$, that  $E<E_c d_s/\sqrt{R_H l_c}\simeq X_s E_c$. Actually,  looking at the detailed results in figure~3 we see that in the non-resonant diffusion  regime (i.e. for $E>0.2 E_c$) the suppression factor starts to become noticeable ($G\simeq 0.9$) for $E\simeq X_sE_c$ and it
reaches the value $G= 0.5$ for the energy $E_{0.5}$ given by
\begin{equation}
E_{0.5}\simeq \frac{1}{4}X_sE_c.
\end{equation}
 Hence,  introducing the corresponding rigidity value $R_{0.5}\equiv E_{0.5}/eZ$
we find the following relation between the different parameters involved (valid for $E_{0.5}>0.2 E_c$)
\begin{equation}
\frac{d_s}{100\ {\rm Mpc}}\frac{B}{{\rm nG}}\sqrt{\frac{l_c}{\rm Mpc}}\simeq \frac{R_{0.5}}{\rm 0.4\ EV}.
\label{e05.eq}
\end{equation}

We note that the expression in eq.~(\ref{jza.eq}) is valid as long as the interaction energy losses are small. In the case of nuclei, if photo-disintegration processes produce some amount of secondaries these particles will diffuse  similarly as their parent nuclei  and their final suppression will then be similar\footnote{Actually the suppression factor $G$ may be slightly larger for secondary nuclei  due to their preferential production at relatively higher redshifts  where Lorentz factors are larger and radiation backgrounds more relevant, but we ignore these effects which are not expected to lead to large differences.}. We will hence apply in the following the suppression factor to the fluxes of both primary and secondary nuclei surviving at low energies.

We have checked that considering proton scenarios with large maximum energies ($E_{max}>10^{20.5}$~eV) we obtain results consistent with the numerical simulations  in \cite{le05} and \cite{be07} (after accounting for the slightly different diffusion coefficients adopted), while considering mixed composition scenarios with large maximum rigidities we obtain results consistent with those in \cite{gl08,al11}.

\section{A  scenario with transition to heavier elements}

We will be interested in the following in scenarios with relatively low maximum rigidities, $E_{max}<(2$--$10)\, Z$~EeV, like those considered in refs.~\cite{al09,al11b}. This is motivated  by the observed trend in the mass composition measured by the Auger Observatory, which suggests a transition towards heavier elements starting above a few EeV, which would naturally result if the lighter elements can no longer be accelerated above the indicated cutoff energies. These `low cutoff scenarios'  have not been studied in detail in the presence of extra-galactic magnetic field  effects (some qualitative considerations were given in \cite{al11b}). Our analytic approximate treatment of diffusion  effects together with the detailed results of CR propagation in the absence of magnetic fields obtained using the CRpropa code \cite{crpropa} allows us to obtain the predictions for the CR spectra and composition in these models.

\begin{figure}
\centerline{\epsfig{width=2.2in,angle=-90,file=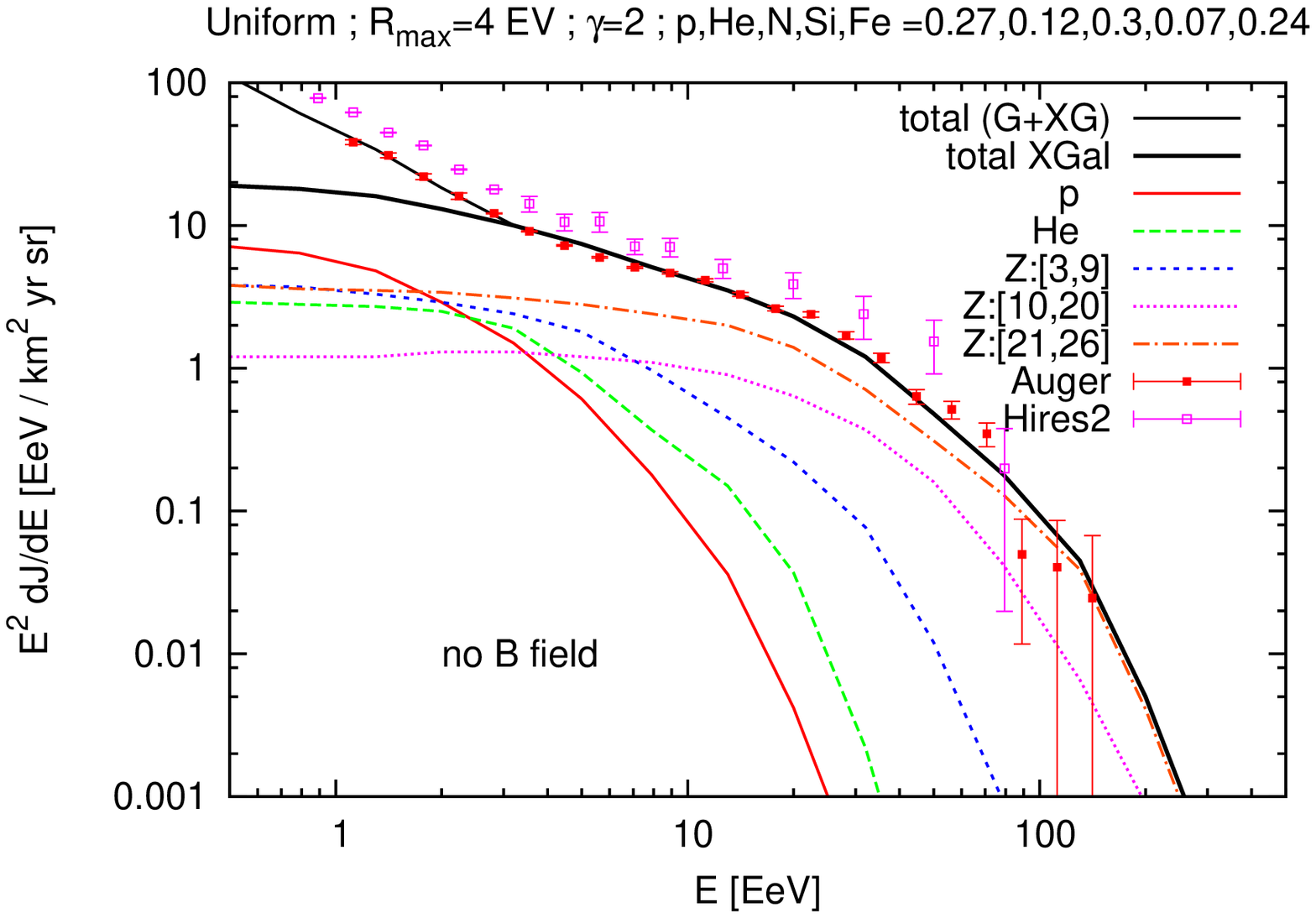}\epsfig{width=2.3in,angle=-90,file=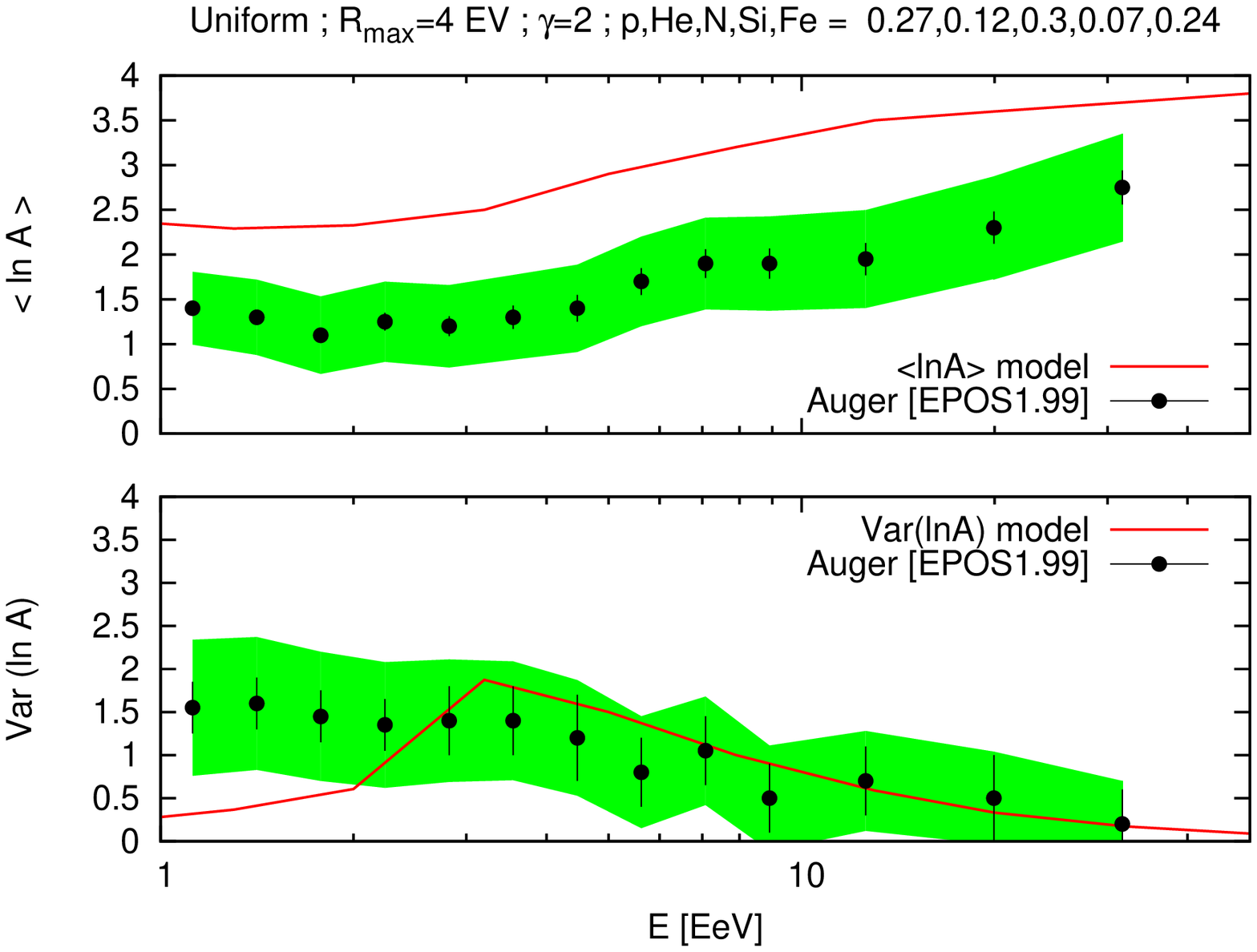}}
\centerline{\epsfig{width=2.2in,angle=-90,file=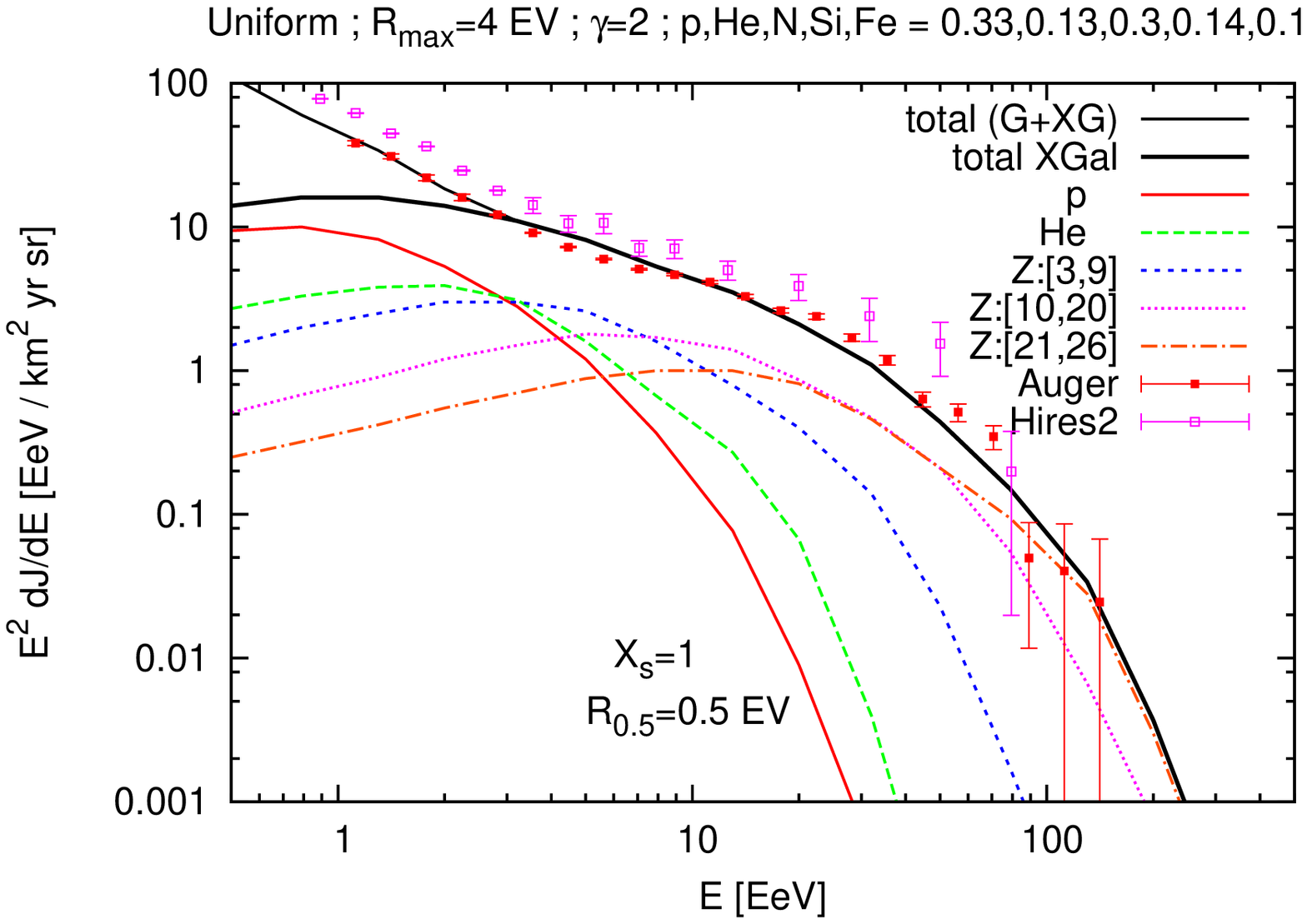}\epsfig{width=2.3in,angle=-90,file=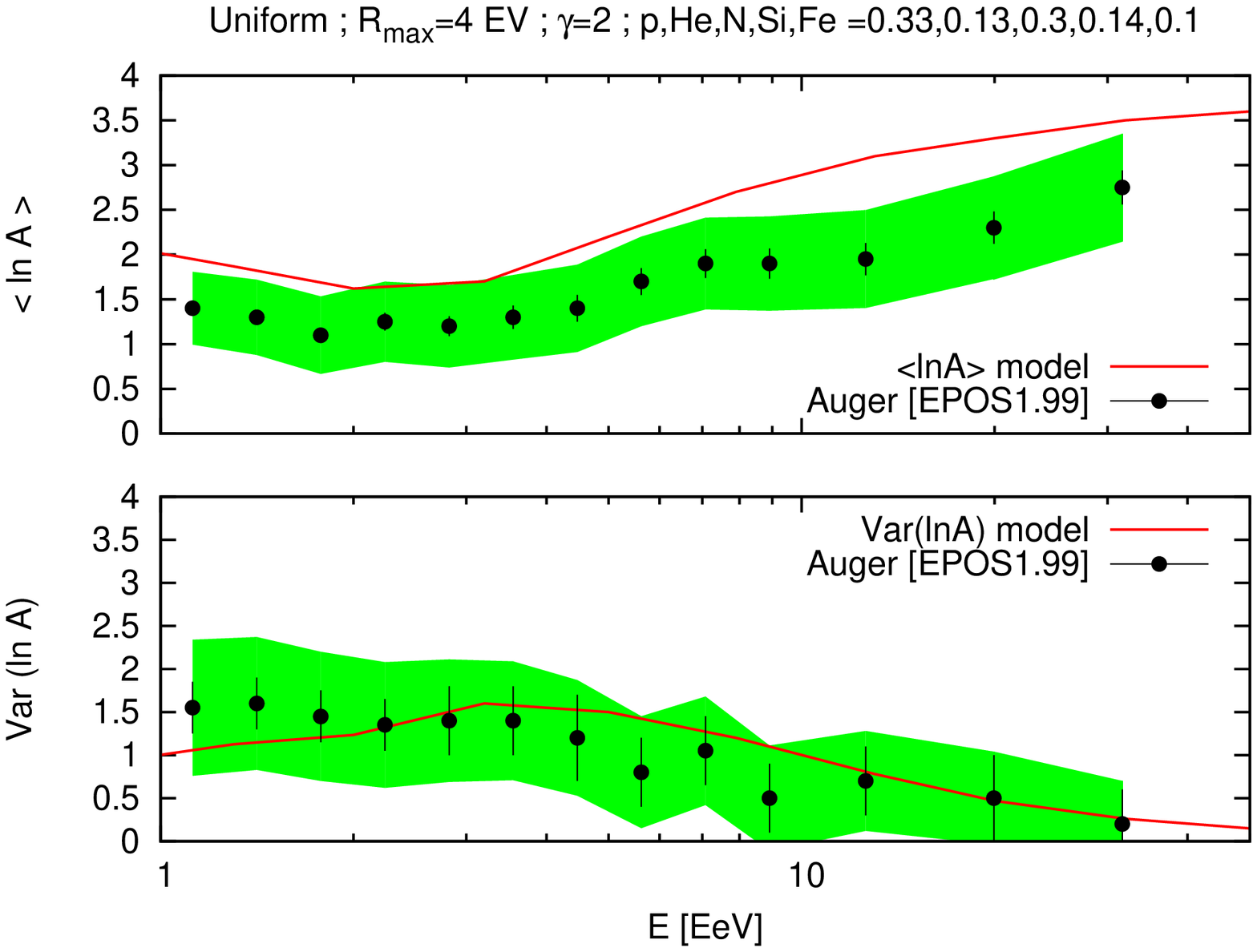}}
\centerline{\epsfig{width=2.2in,angle=-90,file=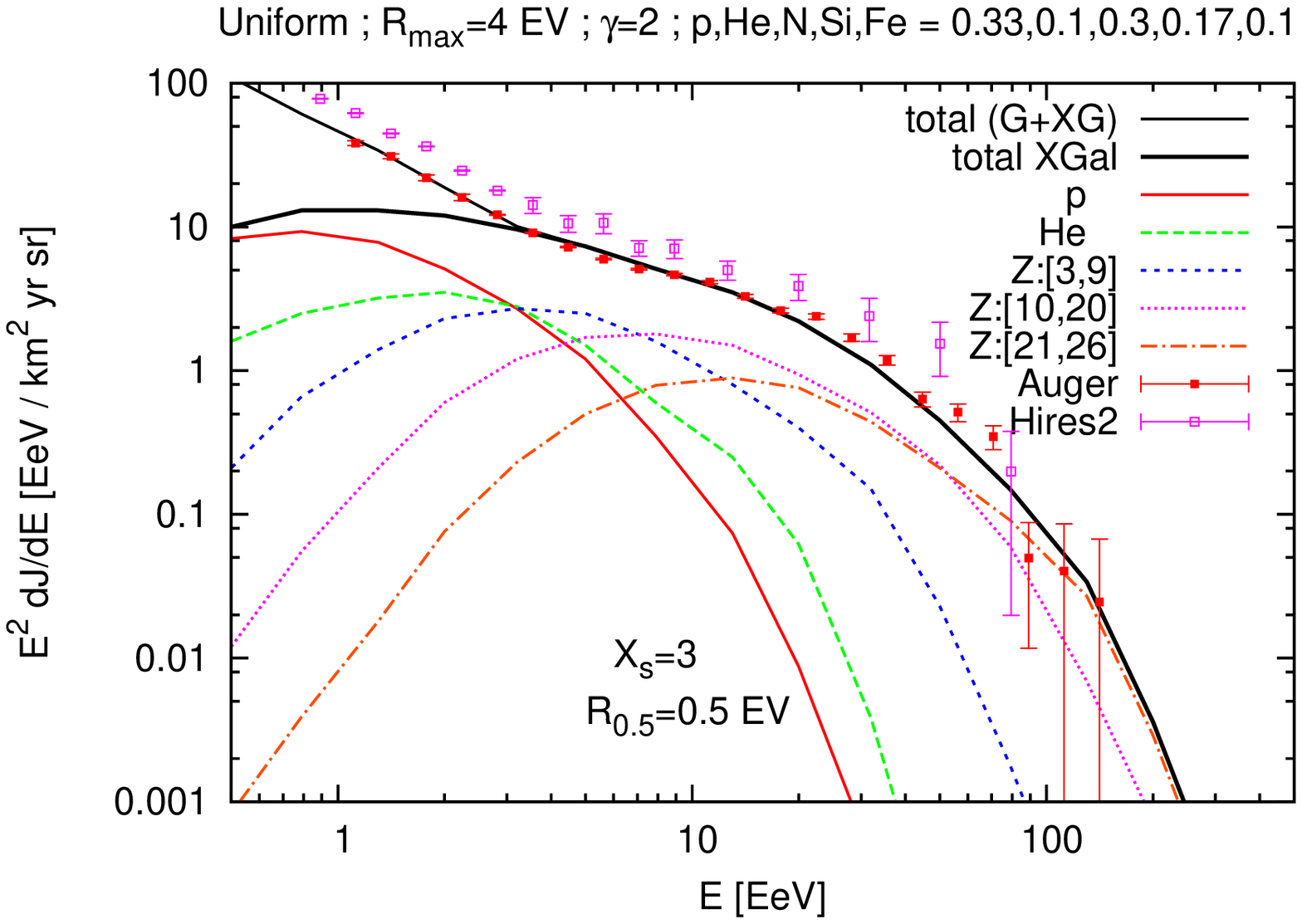}\epsfig{width=2.3in,angle=-90,file=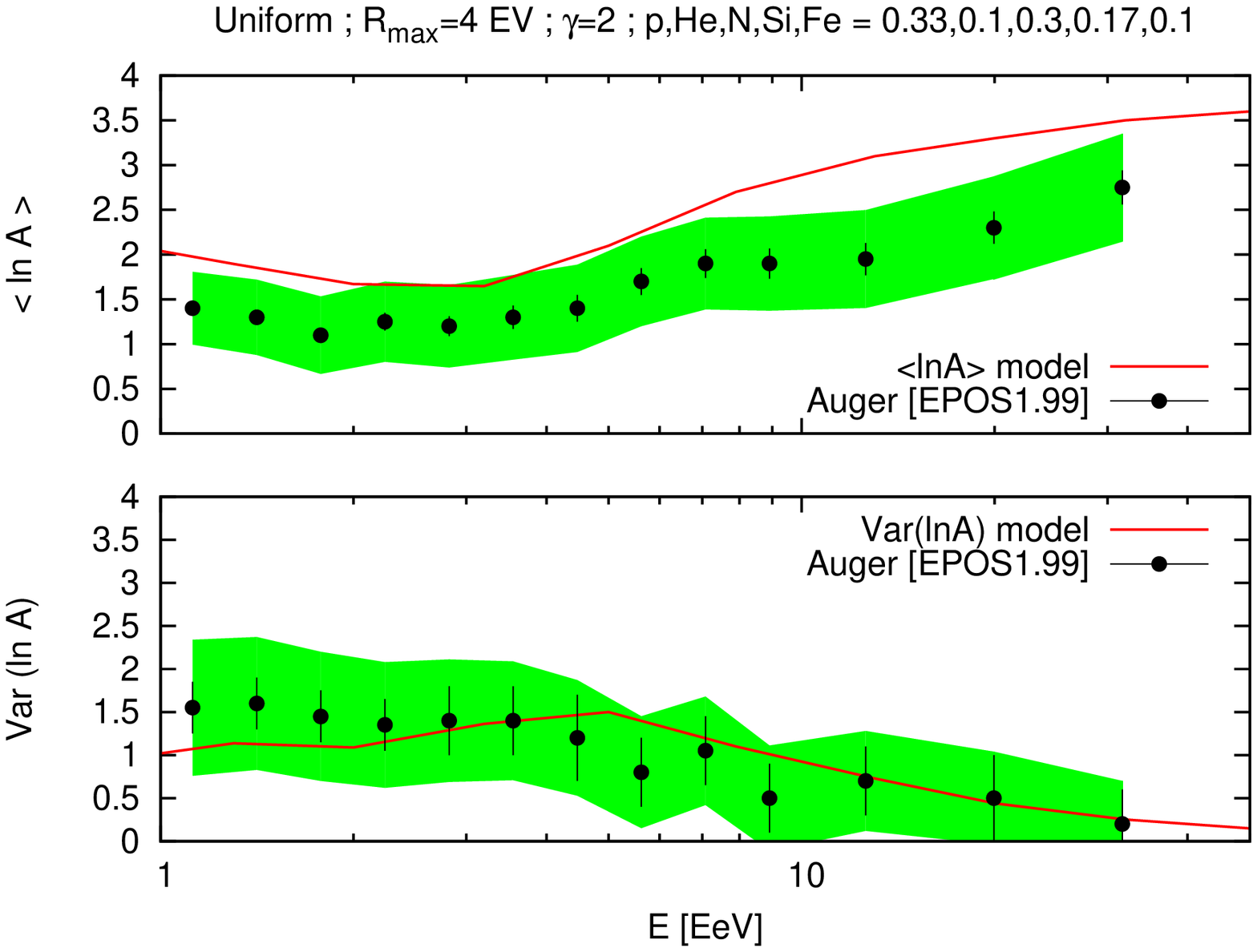}}
\vskip 1.0 truecm
\caption{Uniform source evolution with $\gamma=2$ and $R_{max}=4$~EV for no magnetic field effects (top) and for the case of diffusion with $X_s=1$ (middle) and $X_s=3$ (bottom), adopting $R_{0.5}=0.5$~EV. The individual fractional source compositions are indicated in the corresponding figure labels. Left panels depict the spectrum, while right panels the values of $\langle \ln A\rangle$ and V(ln$A$).} 
\label{fig5}
\end{figure}

As an illustration  we show in fig.~\ref{fig5} the spectra and composition (both the average value of $\ln A$ and its variance, $V(\ln A)$) for scenarios with $E^{-2}$ source spectra  with maximum rigidity $R_{max}\equiv E_{max}/eZ=4$~EV.
We assume that the source spectra consist of
  different fractions of the  representative elements p, He, N, Si and Fe, which are indicated in the figure labels in each case. These fractions are chosen so as to approximately reproduce the observed spectra measured by the Auger Collaboration \cite{augerspec}, which is also shown in the figure together with the one measured by the HiRes  Collaboration \cite{hiresspec}. The three panels correspond to no magnetic field effect (top), diffusion such that $R_{0.5}=0.5$~EV with $X_s=1$ (middle) and $X_s=3$ (bottom). 

The extra-galactic fluxes are obtained using CRpropa and we incorporate the effects of magnetic field diffusion by multiplying the fluxes obtained for each final mass by the corresponding suppression factor $G(E/E_c)$, adopting the expression obtained in eq.~(\ref{gfit.eq}).  The spectra of the different mass groups  as well as the total extra-galactic flux are displayed. 
In addition to the extra-galactic fluxes we also include an additional `galactic' component so as to match the total spectrum down to $\sim 1$~EeV. This extra component may be just the high energy tail of the galactic CRs but may eventually also receive contributions from fainter but closer extra-galactic sources, less suppressed by magnetic diffusion effects.

\begin{figure}[t]
\centerline{\epsfig{width=2.2in,angle=-90,file=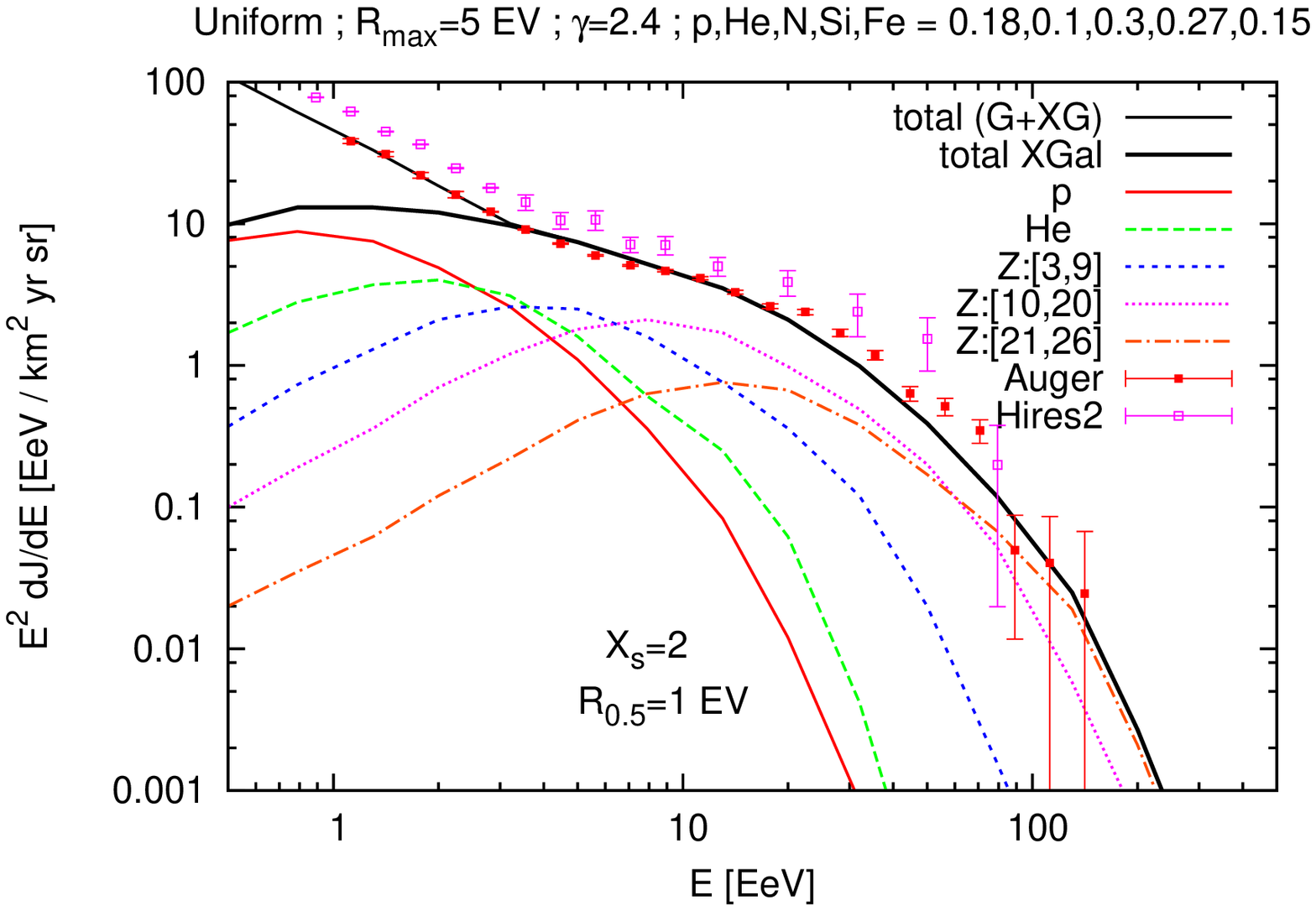}\epsfig{width=2.3in,angle=-90,file=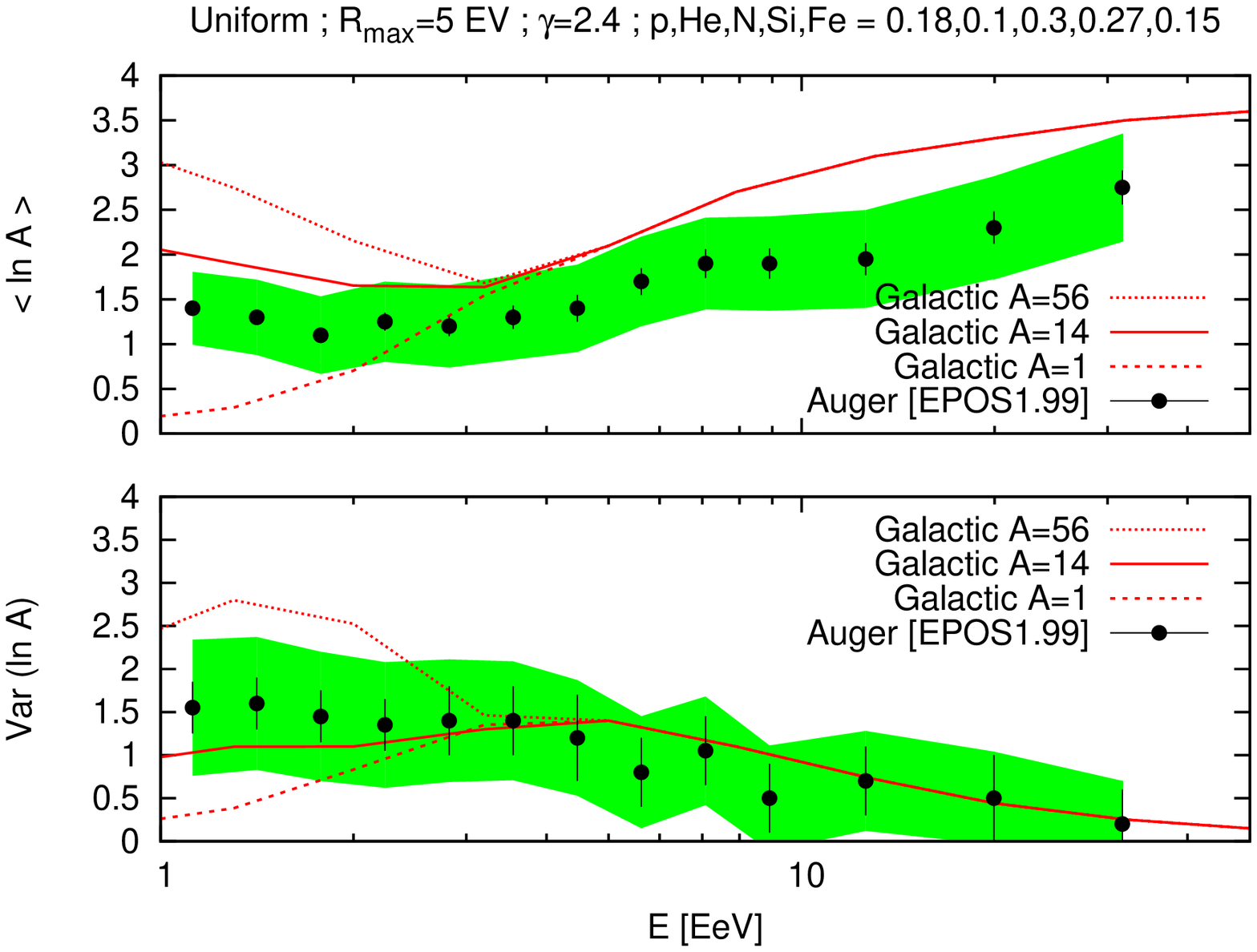}}
\vskip 1.0 truecm
\caption{Uniform source evolution with $\gamma=2.4$, $R_{max}=5$~EV, $X_s=2$ and $R_{0.5}=1$~EV. The composition is shown for different assumptions for the composition of the  `galactic component'. } 
\label{fig6}
\end{figure}

The right panels in the figures show the expected average value of $\ln A$ as well as its variance, and they are compared with the results inferred for these quantities from the measurements of the maximum of the longitudinal development of the showers  and its fluctuations, determined with the fluorescence technique by the Auger Collaboration \cite{augerlna}. The results displayed correspond to the values inferred  adopting the EPOS1.99 hadronic interaction model, with the shaded bands indicating the  associated systematic uncertainties. Somewhat shifted results are obtained  assuming different  hadronic models to interpret the measurements (see  \cite{augerlna}).
We note that below 3~EeV the  contribution from the `galactic' component can be  sizeable and it is relevant for the determination of the resulting composition. The values of $\langle \ln A\rangle$ and $V(\ln A)$ below 3~EeV hence depend on the assumed composition of this extra component. In fig.~\ref{fig5} we just assumed it consisted of $A=14$ nuclei with negligible variance (since this simple assumption led to a rough agreement with the observed values in the cases including magnetic diffusion). We show in fig.~\ref{fig6} how these predictions change for different assumptions on the value of $\ln A$ for the extra component, for the cases of pure Fe, N and p. Had we assumed  a non-negligible intrinsic variance in the mass composition of this extra component, the results for $V(\ln A)$ would have been slightly higher at $E<3$~EeV.  In fig.~\ref{fig6} we adopted $R_{max}=5$~EV and $R_{0.5}=1$~EV, considering  $\gamma=2.4$ to illustrate how a steeper source spectra can also lead to reasonable results.

\begin{figure}[t]
\centerline{\epsfig{width=2.2in,angle=-90,file=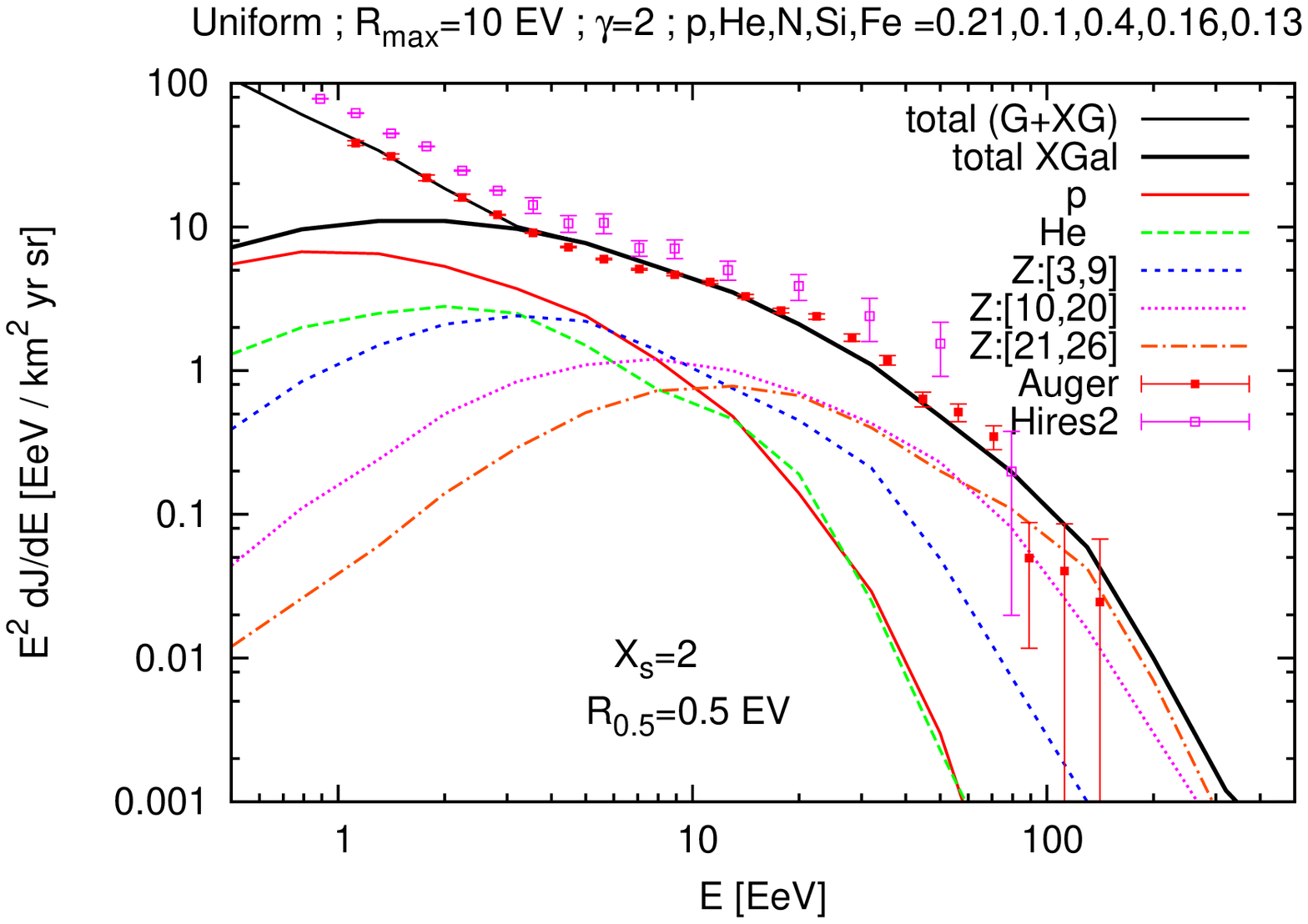}\epsfig{width=2.3in,angle=-90,file=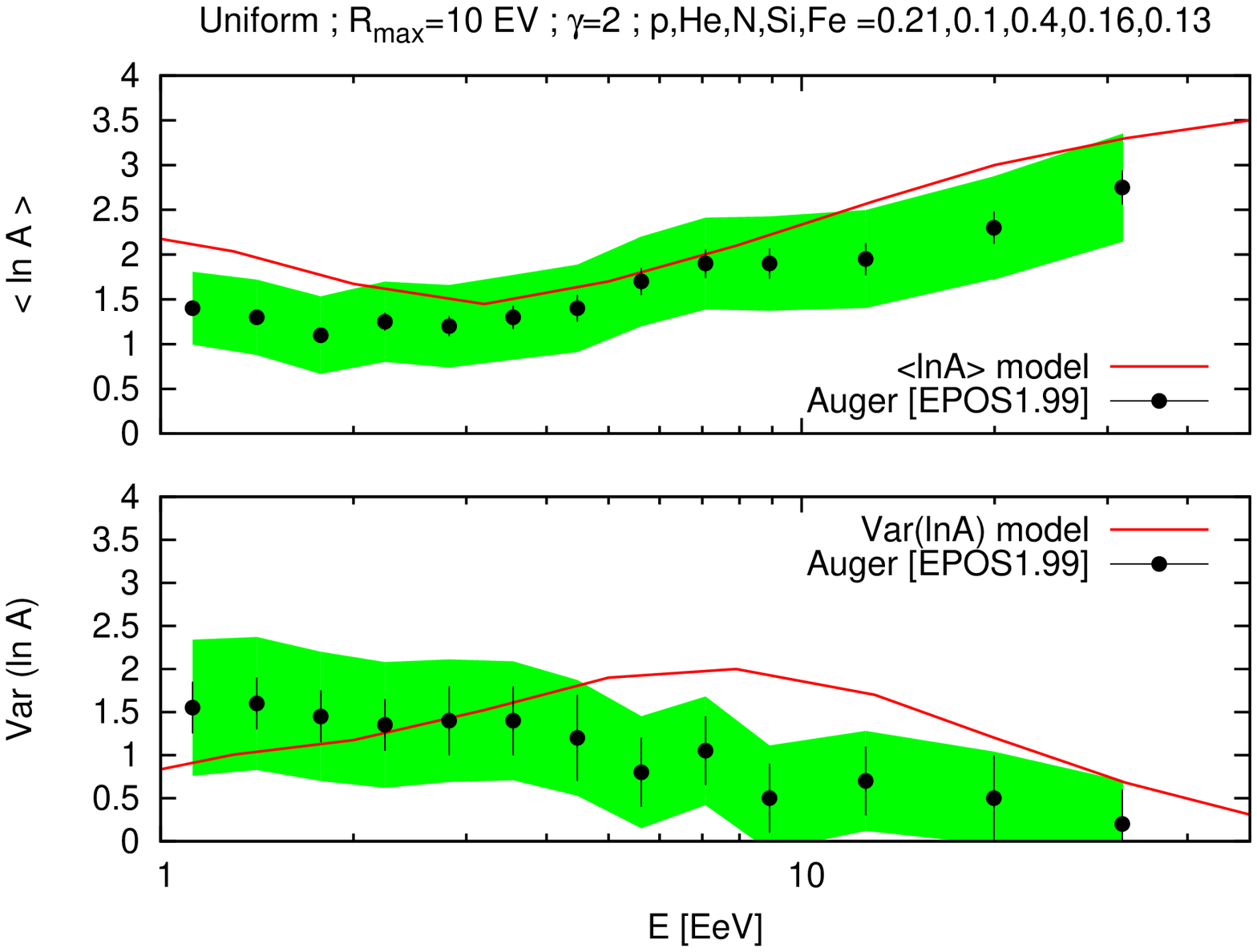}}
\centerline{\epsfig{width=2.2in,angle=-90,file=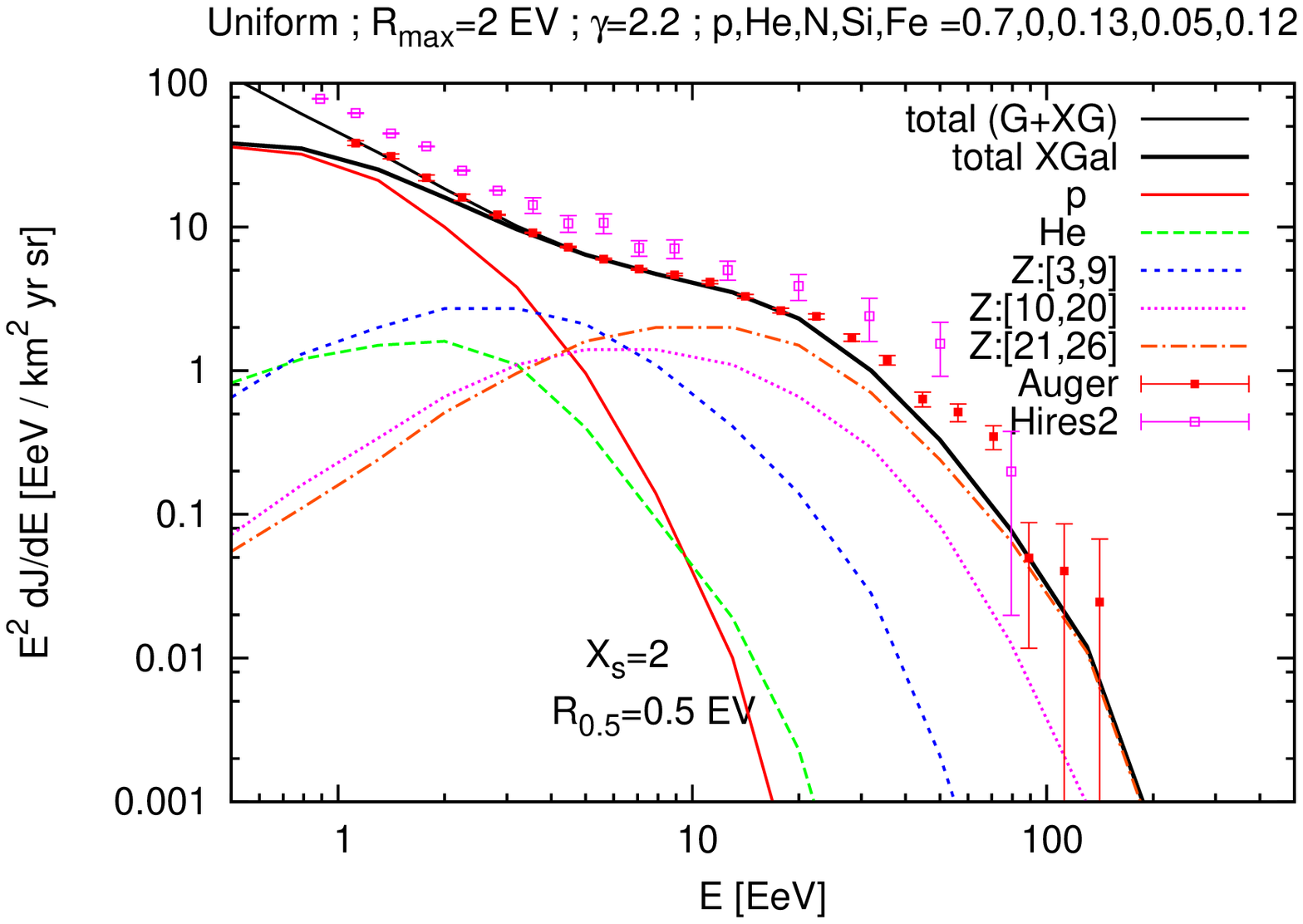}\epsfig{width=2.3in,angle=-90,file=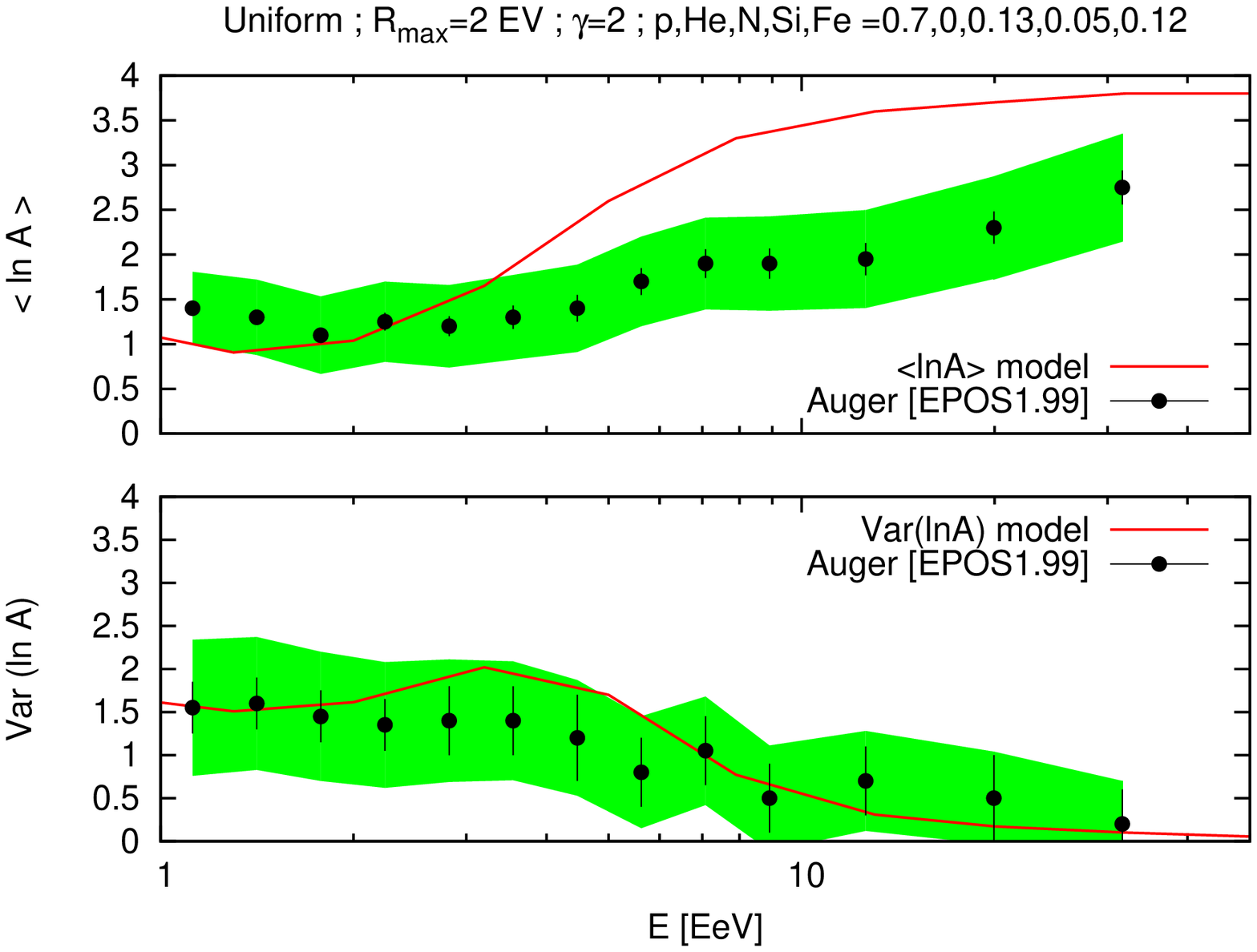}}
\vskip 1.0 truecm
\caption{Uniform source evolution for $X_s=2$ and $R_{0.5}=0.5$~EV. Top panel is for $\gamma=2$ and  $R_{max}=10$~EV while the bottom  panel is for $\gamma=2.2$ and  $R_{max}=2$~EV. } 
\label{fig7}
\end{figure}

\begin{figure}[t]
\centerline{\epsfig{width=2.2in,angle=-90,file=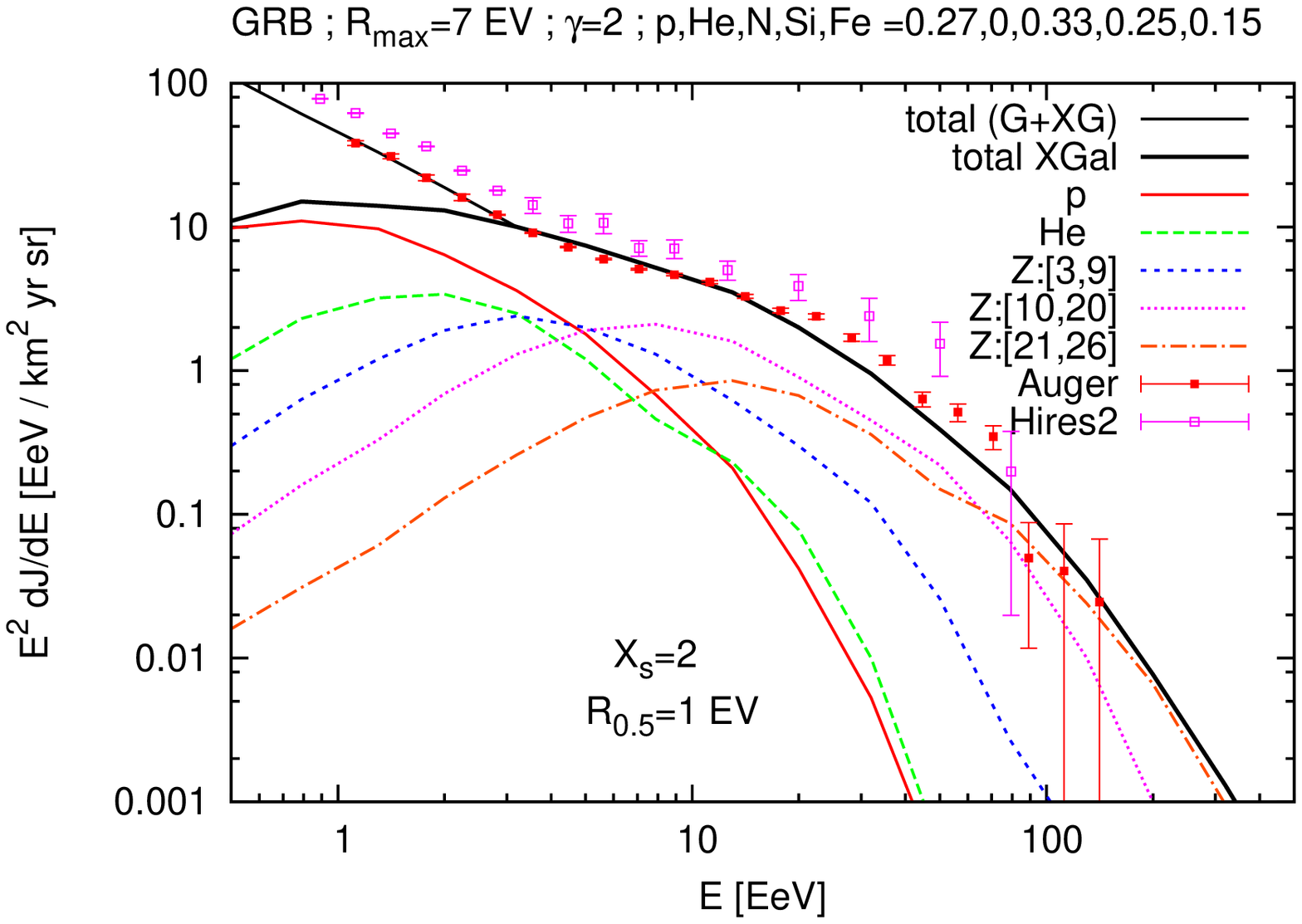}\epsfig{width=2.3in,angle=-90,file=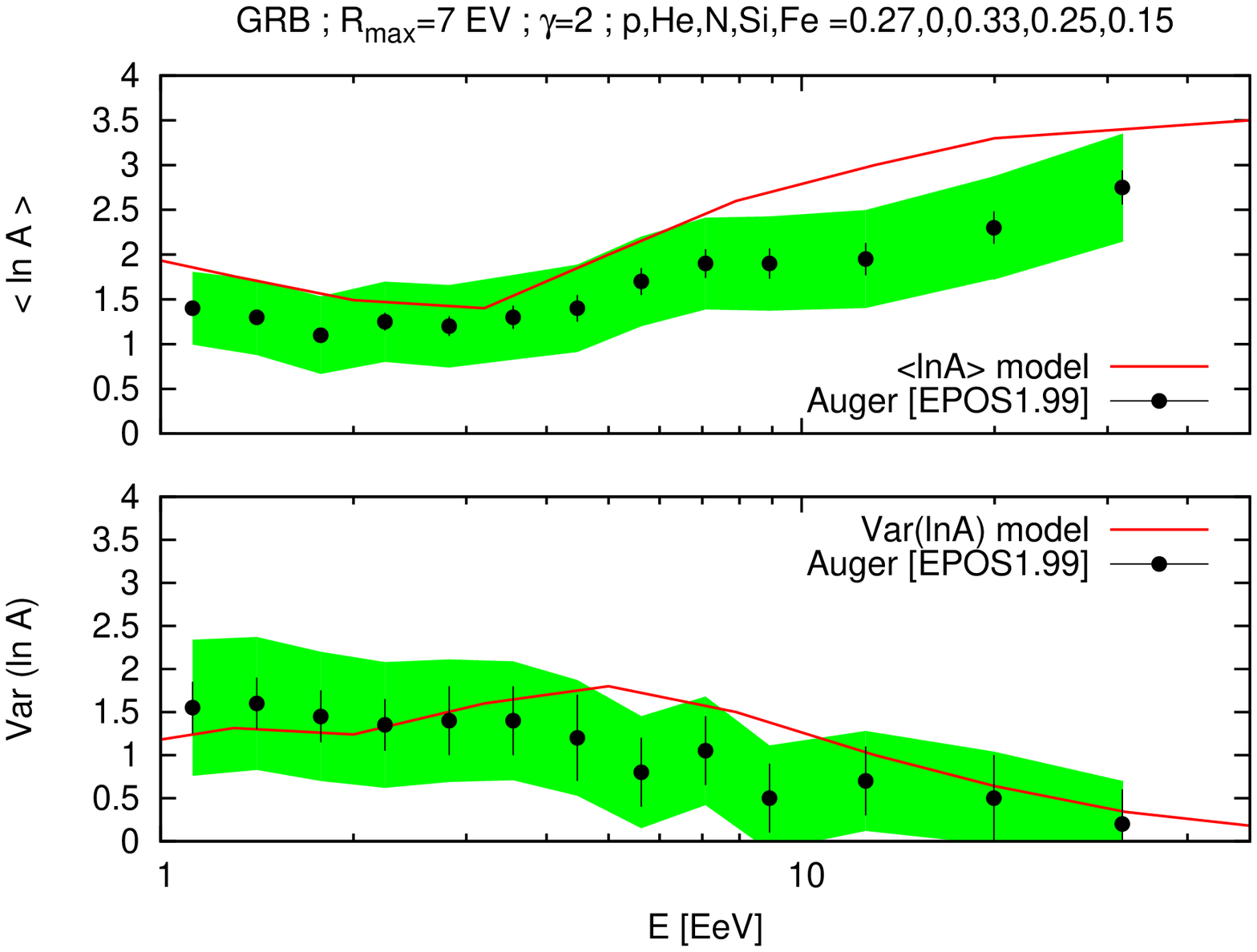}}
\vskip 1.0 truecm
\caption{GRB source evolution with $\gamma=2$, $R_{max}=7$~EV, $X_s=2$ and $R_{0.5}=1$~EV.} 
\label{fig8}
\end{figure}

To illustrate the effect of the maximum rigidity cutoff adopted we show in 
fig.~\ref{fig7}  the results for $R_{max}=10$~EV,  with $\gamma=2$ (top panel), and for $R_{max}=2$~EV, with $\gamma=2.2$ (lower panel). The first one  shows the tendency to increase the variance of $\ln A$ when the light elements contribute up to higher energies, while the second one shows the tendency to increase $\langle \ln A\rangle$ when the cutoff is lowered.  
Finally, fig.~\ref{fig8} exemplifies the effect of the source luminosity evolution. It displays the results for the case of strong evolution of the source  emissivity, i.e. with $f(z)$ following the Gamma Ray Burst rate (corresponding to the SFR6 evolution obtained in \cite{le07}), adopting $\gamma=2$ and $R_{max}=7$~EV. In this case the extragalactic proton component can be more enhanced below the ankle.

We should mention that the  finite distance to the closest sources can also make the `GZK' suppression at the highest energies stronger than in the case of a continuous source distribution (see \cite{be07,gl08,ta11}). However, this effect is expected to be more pronounced in scenarios with high rigidity cutoffs for which the effects of the `GZK' suppression are important, while in the scenarios we considered with low rigidity cutoffs this should have only a minor impact, and the high energy suppression of the spectrum is largely due to the source acceleration cutoff itself rather than to the propagation effects. Given the arbitrariness in the shape of the cutoff function adopted, which ultimately models the suppression at the highest energies in our case, we  do not include the additional suppression from the finite source distance effect at the highest energies.

In the previous plots of the spectrum and composition (figs.~\ref{fig5} to \ref{fig8}) we fixed the parameters $X_s$ and $R_{0.5}$, which determine the shape and location respectively of the suppression due to the magnetic diffusion. These parameters are ultimately related to the extra-galactic turbulent magnetic field
strength $B$ and coherence length $l_c$ as well as to the typical separation between sources $d_s$ (and hence to $n_s$), through eqs.(\ref{xs.eq}) and (\ref{e05.eq}). However, the relation between these quantities is not unique, so that different combinations of $B$, $l_c$ and $d_s$ values can give rise to the same results. For instance, to have $R_{0.5}=0.5$~EV and $X_s=1$ would require that $d_s B \sqrt{l_c}\simeq 125\ {\rm nG}\,{\rm Mpc}^{3/2}$ and $d_s/\sqrt{l_c}\simeq 65 \sqrt{\rm Mpc}$  (which can be obtained for instance with  $d_s\simeq 50$~Mpc, $l_c\simeq 0.6$~Mpc and  $B\simeq 3$~nG). On the other hand, if we consider a given value for $R_{0.5}$ one finds that for $X_s>2$ the shape of the suppression is essentially independent of $X_s$ so that the resulting spectra and composition will be quite independent of the actual  value of $X_s$, with this parameter just determining  the value for $d_s/\sqrt{l_c}$  inferred.

In the previous plots we did not attempt to obtain detailed fits to the Auger results but just tried to show that the diffusion in extra-galactic magnetic fields can help to give rise to the observed trends in $\langle\ln A\rangle$ and $V(\ln A)$ in scenarios compatible with Fermi acceleration (while fits to Auger results ignoring magnetic effects tend to prefer $\gamma<2$ \cite{al09,ho10,al11,ta11}). Notice also that a detailed interpretation of
 the composition results depends actually on the assumed hadronic models. In addition, one should keep in mind that results from HiRes and Telescope Array are compatible with a light component even at energies above 10~EeV \cite{hiresx,tax}. Moreover, the spectrum is still affected by the systematics of the energy determination, as is evident from the differences between Auger and HiRes spectra.

\section{Conclusions}
In this work we have considered in detail the low energy spectral suppression appearing when the density of extra-galactic UHECR sources is low  and CRs diffuse in turbulent extra-galactic magnetic fields. We obtained simple fits to the flux suppression factor valid in the limit in which adiabatic losses dominate over interaction ones, which holds for $E<Z$~EeV. Scenarios in which the flux suppression
appear in this energy regime are particularly relevant given the indications in favor of the presence of a significant light extra-galactic component at EeV energies. On the other hand, we considered scenarios with low maximum rigidities, $R_{max}=(2$--10)~EV, in which a transition to heavier elements for increasing energies naturally occurs. The magnetic diffusion effects then allow to suppress the extragalactic nuclei contribution at $E<Z$~EeV without requiring the introduction of very hard source spectra and can hence lead to results more consistent with the overall composition trends measured by the Auger Collaboration also for source spectra compatible with Fermi shock acceleration ($\gamma=2$--2.4).

\section*{Acknowledgments}
We are grateful to D. Harari for useful discussions. We thank the authors of CRpropa for making their code public. 
 Work supported by CONICET and ANPCyT, Argentina.

\section*{Appendix}
In this Appendix we discuss the low energy attenuation of the proton spectrum in the case in which the energy losses due to pair creation off CMB photons are also included. This allows  to illustrate their impact and shows the applicability of the results obtained with adiabatic losses alone.

We include pair creation losses off CMB photons following \cite{ch92}, from which  we compute the energy loss coefficient $\left. b\equiv -{\rm d}E/{\rm d}t\right|_{int}$ due to interactions alone. Taking into account the redshift evolution of the CMB photon energy and of its  density, one has that
\begin{equation}
b(E,z)=(1+z)^2b_0[E(1+z)],
\end{equation}
with $b_0$ the value computed at $z=0$. The original energy at redshift $z$, $E_g(E,z)$, is then obtained by integrating the equation
\begin{equation}
\frac{{\rm d}E}{{\rm d}z}=\frac{E}{1+z}+\frac{(1+z)b_0[(1+z)E]}{H_0\sqrt{\Omega_m(1+z)^3+\Omega_\Lambda}},
\end{equation}
which accounts for both adiabatic and pair production losses.
Another relation which is necessary to compute $J_s$ in eq.~(\ref{js.eq}) is (see Appendix in \cite{be06b})
\begin{equation}
\frac{{\rm d}E_g}{{\rm d}E}=(1+z)\exp\left[\frac{1}{H_0}\int_0^z {\rm d}z'\,\frac{(1+z')^2}{\sqrt{\Omega_m(1+z)^3+\Omega_\Lambda}} \left.\frac{{\rm d}b_0(E')}{{\rm d}E'}\right|_{E'=(1+z')E_g}\right],
\end{equation}
where to a good approximation one can use that d$b_0/{\rm d}E\simeq b_0/E$ \cite{be06b}.

\begin{figure}[t]
\centerline{\epsfig{width=2.8in,angle=0,file=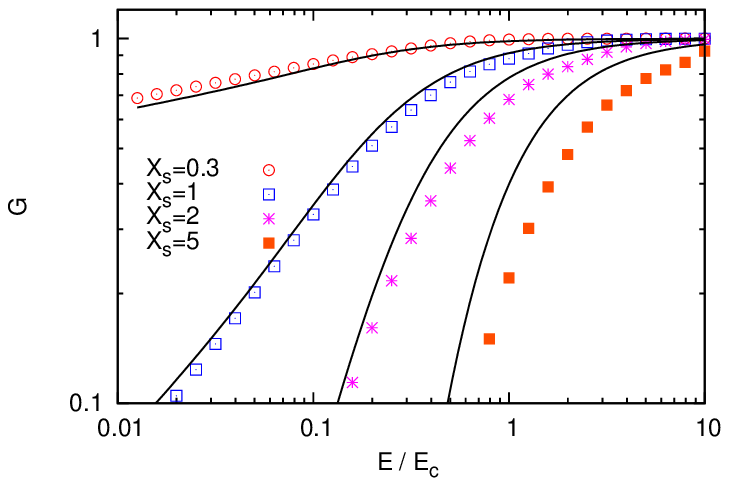}\epsfig{width=2.8in,angle=0,file=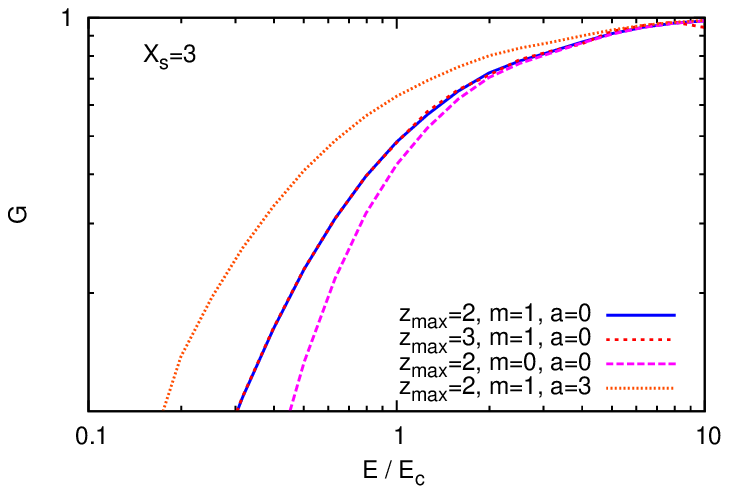}}
\vskip 1.0 truecm
\caption{Suppression factor vs. $E/E_c$ adopting $\gamma=2$, $B=1$~nG, $l_c=1$~Mpc and  $E_{max}=10$~EeV. Left panel  adopts a uniform source evolution and different values of $X_s$  (solid lines indicate the fits obtained ignoring pair creation losses, see fig.~\ref{fig3}). Right panel is for $X_s=3$ and different cosmological evolution parameters, similar as fig.~\ref{fig4} but including pair losses. } 
\label{fig9}
\end{figure}

Combining all this we can then compute the suppression factor including energy losses due to pair production, with the results shown in fig.~\ref{fig9}.  We here adopted $B=1$~nG, $l_c=1$~Mpc (leading to $E_c\simeq 0.9$~EeV) and $E_{max}=10$~EeV (the results have only a mild dependence on the value of $E_{max}$ adopted).
The left panel is the equivalent to fig.~\ref{fig3} while the right panel to  fig.~\ref{fig4}.
 As can be seen from the left panel the resulting suppression factors are essentially unchanged  at low energies (solid lines represent the fits obtained ignoring pair creation losses which were also shown in fig.~\ref{fig3}).  Only in the cases in which $E_{0.5}\geq 1$~EeV, corresponding to large values of $X_s$ (actually for $X_sE_c\geq 4$~EeV)
the flux suppression has some noticeable changes, as in the case shown for $X_s=5$, and to a minor extent for the case $X_s=2$. For large $X_s$ the suppression is approximately similar in shape to that obtained ignoring pair losses  but it is slightly shifted to higher energies. This also implies that the final spectra obtained assuming a given value of $E_{0.5}$ will be quite independent of whether pair production losses are included or not, but eventually what can change for large values of $X_s$ is the relation between $E_{0.5}$ and the underlying magnetic field parameters and source distances (tending to reduce the required values of $B$).

We note that when pair losses are included one finds that for $E>1$~EeV and  at high redshifts the initial energy $E_g(E,z)$ can be much larger than $E(1+z)$, showing an explosive increase with redshift as soon as $E_g>2$~EeV. When this happens, the contribution to $J_s$ arising from higher redshifts gets very suppressed. This explains the behavior observed, where the increased flux suppression is the result of the competition between two opposite effects. On one side the larger values of $E_g(E,z)$ obtained including pair losses lead to larger values for $\lambda$, what would suggest that the flux should be less suppressed (see eq.~(\ref{ffit.eq})). However, the fact that higher redshifts contribute less to $J_s$ implies that the overall suppression, caused by the fact that CRs from nearby sources have not enough time to diffuse from small redshifts, ends up being stronger.  The behavior of $E_g(E,z)$   also explains why including pair losses the suppression factors become quite insensitive to the maximum redshift considered, as is apparent from the right panel in fig.~\ref{fig9}.

We hence conclude that  when pair creation losses are included the suppression factors for given values of $X_s$ and $E_{0.5}$ remain essentially unchanged as long as $E_{0.5}<1$~EeV.
The case of nuclei should be similar (with the suppression factors below $Z$~EeV remaining unaffected) but the inclusion of photo-disintegration effects would make the analysis more involved in this case.

\end{document}